\documentclass{appolb}
\usepackage{graphicx}

\begin{document}
\title{MESON ASSISTED DIBARYONS\footnote{Presented at the Jagiellonian 
Symposium on Fundamental and Applied Subatomic Physics, Cracow, June 2015.}} 
\author{Avraham Gal\address{Racah Institute of Physics, The Hebrew University, 
Jerusalem 91904, Israel}} 

\maketitle

\begin{abstract}
We discuss a new type of $L=0$ positive-parity dibaryons, $\pi BB'$, where 
the dominant binding mechanism is provided by resonating $p$-wave pion-baryon 
interactions. Recent calculations of such pion assisted dibaryons are reviewed 
with special emphasis placed on the non-strange $I(J^P)$=$1(2^+)$ $N\Delta$ 
dibaryon ${\cal D}_{12}(2150)$ studied recently at JLab, and on the $0(3^+)$ 
$\Delta\Delta$ dibaryon ${\cal D}_{03}(2380)$ discovered recently by 
the WASA-at-COSY Collaboration. We discuss recent searches by the HADES 
Collaboration at GSI and by the E15 and E27 Experiments at J-PARC for 
a strangeness $\cal S$=$-$1 $I(J^P)$=$\frac{1}{2}(0^-)$ $K^{-}pp$ dibaryon 
and perhaps also for a strange $I(J^P)$=$\frac{3}{2}(2^+)$ $N\Sigma(1385)$ 
pion assisted dibaryon ${\cal Y}_{\frac{3}{2}2}(2270)$. Charm $\cal C$=$+$1 
dibaryons, predicted with these same $I(J^P)$ values, are also briefly 
reviewed. 
\end{abstract}

\section{Introduction} 
\label{sec:int} 

The present overview is focused on the notion of pion assisted dibaryons, 
$\pi BB^\prime$. The idea is to enhance the binding of $L=0$ $BB^\prime$ 
configurations through the strong $p$-wave $\pi B$ and $\pi B'$ attraction.
In the ${\cal S}=0$ non-strange sector, for the $\pi NN$ system, we show 
how certain $N\Delta$ near-threshold quasibound states emerge, and for 
the $\pi N\Delta$ system we show how certain $\Delta\Delta$ quasibound 
states emerge, notably the $I(J^P)$=$0(3^+)$ ${\cal D}_{03}(2380)$ 
dibaryon discovered recently by the WASA-at-COSY 
Collaboration \cite{wasa11,wasa13,wasa14}. 

In the strangeness ${\cal S}=-1$ sector, we focus attention to 
a $\pi\Lambda N-\pi\Sigma N$ dibaryon in a spin and isospin stretched 
configuration $I(J^P)$=$\frac{3}{2}(2^+)$ predicted near the $\pi \Sigma N$ 
threshold at $\sqrt{s}\approx 2270$~MeV \cite{gg13}.{\footnote{Earlier 
versions of this work are detailed in Refs.~\cite{gg08,gg10}.}} 
This pion assisted dibaryon, resembling a two-body quasibound state of 
$N\Sigma(1385)$ and to a lesser extent $\Delta(1232)Y$, with $Y\equiv \Lambda,
\Sigma$, may be looked for in the same production reactions used to search 
for a $K^-pp$ $I(J^P)$=$\frac{1}{2}(0^-)$ $\bar K$ assisted dibaryon (but 
with $s$-wave $K^-$ meson) which may also be viewed as a $N\Lambda(1405)$ 
quasibound state \cite{oka11}. For a recent overview of $K^-pp$ and its 
implications to $\bar K$--nuclear phenomenology, see Ref.~\cite{gal13}. 
 
In the charm ${\cal C}=+1$ sector, we briefly review two recently suggested 
charmed dibaryons, with $I(J^P)=\frac{1}{2}(0^-)$ \& $\frac{3}{2}(2^+)$ 
configurations, in perfect analogy to the ${\cal S}=-1$ dibaryons discussed 
above. 

\begin{figure}[htb] 
\begin{center} 
\includegraphics[width=0.48\textwidth,height=5cm]{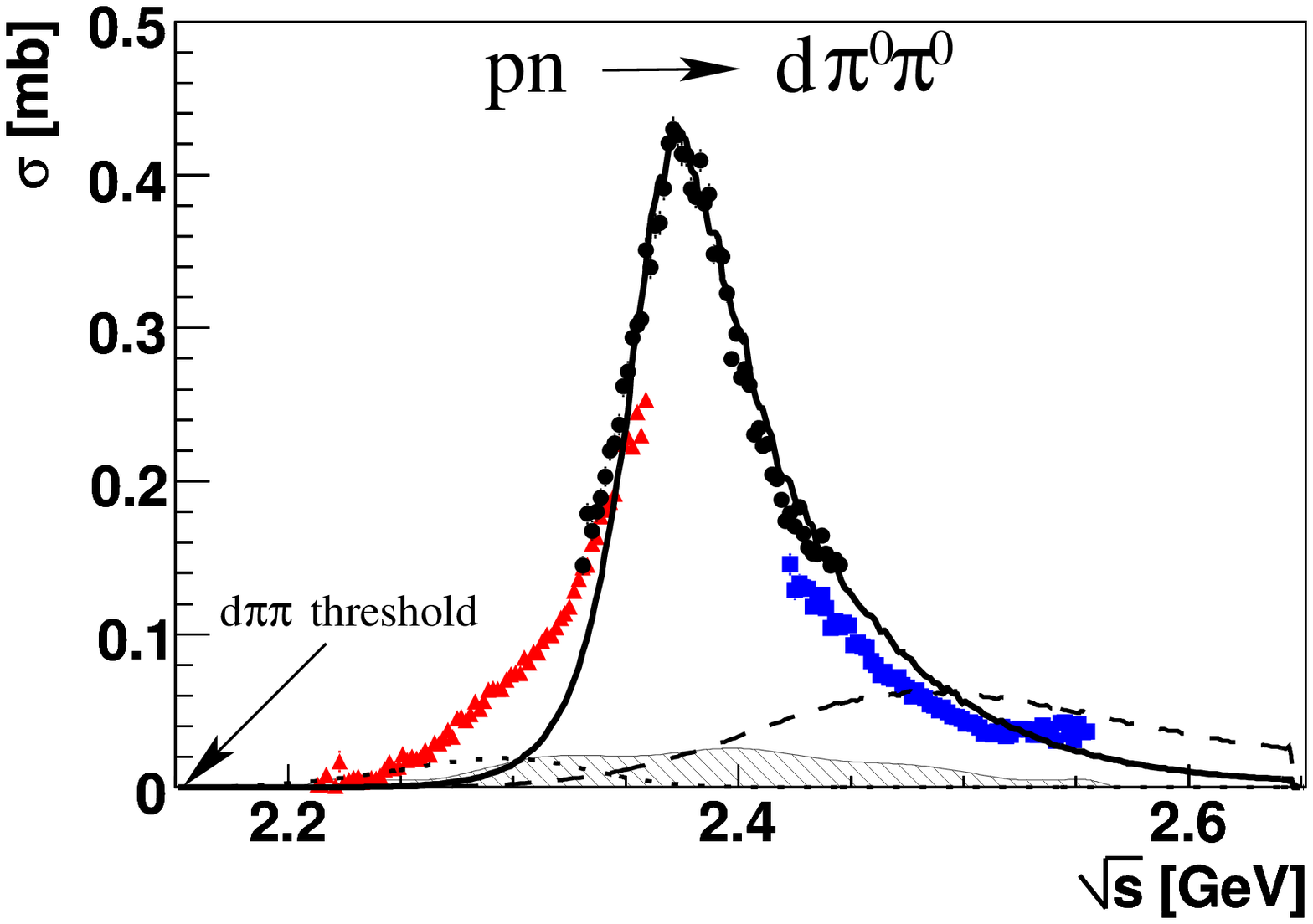} 
\includegraphics[width=0.48\textwidth,height=5cm]{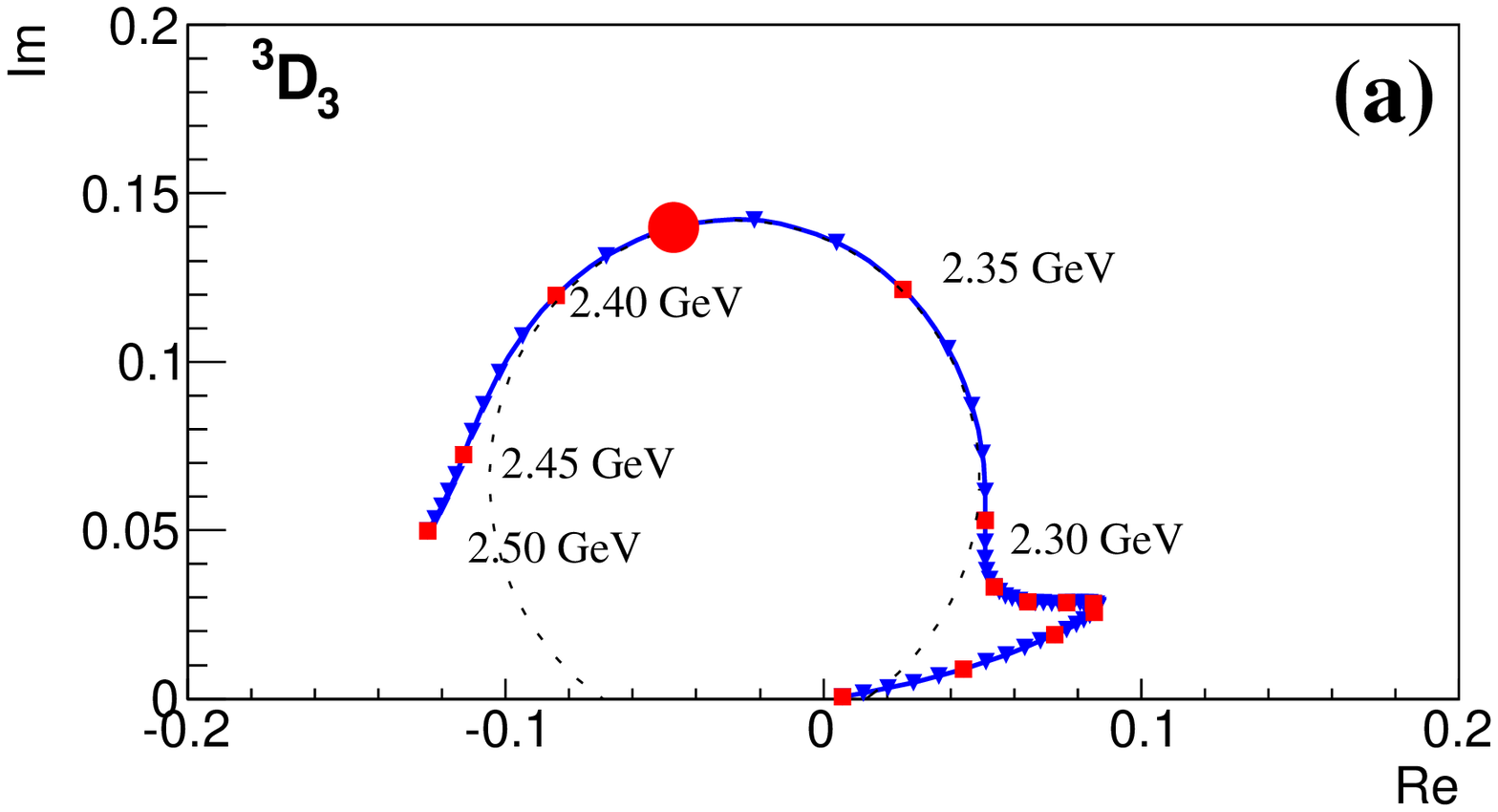} 
\caption{${\cal D}_{03}(2380)$ $\Delta\Delta$ dibaryon resonance signatures 
in recent experiments by the WASA-at-COSY Collaboration. Left: from observing 
a peak in the $pn\to d\pi^0\pi^0$ reaction \cite{wasa11}. Right: from the 
Argand diagram of the $^3D_3$ partial wave in $pn$ scattering \cite{wasa14}.} 
\label{fig:WASA} 
\end{center} 
\end{figure} 

The present overview updates a review of dibaryons published a few years 
ago \cite{gal11} when the mere observation of just a peak in the $pn\to 
d\pi^0\pi^0$ reaction \cite{wasa11}, see left panel of Fig.~\ref{fig:WASA}, 
was not generally accepted as evidence for the $I(J^P)$=$0(3^+)$ 
${\cal D}_{03}(2380)$ $\Delta\Delta$ dibaryon resonance. A corresponding peak 
was subsequently seen also in $pn\to d\pi^+\pi^-$ \cite{wasa13}, with a cross 
section related to that of $pn\to d\pi^0\pi^0$ by assuming an underlying 
${\cal D}_{03}(2380)$ dibaryon resonance. Recent measurements by WASA-at-COSY 
\cite{wasa14} of $pn$ scattering and analyzing power, as shown by the $pn$ 
$^3D_3$ partial wave Argand diagram in the right panel of Fig.~\ref{fig:WASA}, 
provide a `smoking gun' for this dibaryon which is the {\it only} dibaryon 
established unambiguously so far. My own work with Garcilazo, interpreting 
${\cal D}_{03}(2380)$ as a $\pi N\Delta$ pion assisted dibaryon, took a while 
to develop \cite{galgar13,galgar14}. Before getting to this main subject, 
we start in the next section with a brief overview of dibaryon expectations 
from quark models, then moving on to discuss meson assisted dibaryons in the 
non-strange, strange and charmed sectors mentioned above.

\section{Quark models} 
\label{sec:QM}

Historically, discussions of six-quark ($6q$) dibaryons were based on symmetry 
considerations related to the color-magnetic (CM) gluon exchange interaction
\begin{equation} 
V_{CM}=\sum_{i<j}-(\lambda_i\cdot\lambda_j)(s_i\cdot s_j)v(r_{ij}),
\label{eq:cm}
\end{equation}
where $\lambda_i$ and $s_i$ are the color and spin operators of the $i$-th 
quark and $v(r_{ij})$ is a flavor-conserving short-ranged interaction between 
quarks $i,j$. For $L=0$ spatially symmetric color-singlet $n$-quark cluster, 
the matrix element of $v(r_{ij})$ is independent of the particular $i,j$ pair 
and is denoted ${\cal M}_0$, allowing for a closed form summation over $i$ 
and $j$ in Eq.~(\ref{eq:cm}) and resulting in 
\begin{equation}
\langle V_{\rm CM} \rangle = [-\frac{n(10-n)}{4}+\Delta{\cal P}_{\rm f}+
\frac{S(S+1)}{3}]{\cal{M}}_0,
\label{eq:CM}
\end{equation}
where ${\cal P}_{\rm f}$ sums over $\pm 1$ for any symmetric/antisymmetric
flavor pair, $\Delta{\cal P}_{\rm f}$ means with respect to the
SU(3)$_{\rm f}$ $\bf 1$ antisymmetric representation of $n$ quarks, $n=3$
for baryons and $n=6$ for dibaryons, $S$ is the total Pauli spin, and where
${\cal M}_0\sim 75$~MeV from the $\Delta$--$N$ mass difference. The leading 
strangeness ${\cal S}=0,-1,-2,-3$ dibaryon candidates arising from these CM 
considerations are listed in Table~\ref{tab:oka} following Ref.~\cite{oka88}, 
where $\Delta \langle V_{\rm CM} \rangle = \langle V_{\rm CM} {\rangle}_{6q} 
-\langle V_{\rm CM}{\rangle}_{B} -\langle V_{\rm CM} {\rangle}_{B^\prime}$ 
stands for the CM interaction gain in the $6q$ dibaryon configuration with 
respect to the sum of CM contributions from the separate $B$ and $B^{\prime}$ 
$3q$ baryons that define the lowest $BB^{\prime}$ threshold. 

\begin{table}[hbt]
\begin{center}
\caption{Leading $6q$ $L=0$ dibaryon candidates \cite{oka88}, their
$BB^\prime$ structure and the CM interaction gain with respect of the lowest
$BB^\prime$ threshold calculated by means of Eq.~(\ref{eq:CM}). Asterisks 
are used for the ${\bf 10}_{\rm f}$ baryons $\Sigma^{\ast}\equiv\Sigma(1385)$ 
and $\Xi^{\ast}\equiv\Xi(1530)$. The symbol [i,j,k] stands for the Young 
tablaux of the SU(3)$_{\rm f}$ representation, with i arrays in the first 
row, j arrays in the second row and k arrays in the third row, from which 
${\cal P}_{\rm f}$ is evaluated. The ${\overline{\bf 10}}$ SU(3)$_{\rm f}$ 
representation is denoted here ${\bf 10}^{\ast}$.}
\label{tab:oka}
\begin{tabular}{clcccc}
\hline 
\( -{\cal S} \)& SU(3)$_{\rm f}$ & $I$ & $J^{\pi}$ & $BB^\prime$ structure &
$\frac{\Delta \langle V_{\rm CM} \rangle}{M_0}$  \\ 
\hline 
0 & [3,3,0] ${\bf 10}^{\ast}$ & 0 & 3$^+$ & $\Delta\Delta$ & $\,0$  \\ 
1 & [3,2,1] $\bf 8$ & 1/2 & 2$^+$ &
$\frac{1}{\sqrt{5}}(N\Sigma^{\ast}+2\Delta\Sigma)$ & $-1$  \\
2 & [2,2,2] $\bf 1$ & 0 & 0$^+$ &
$\frac{1}{\sqrt{8}}(\Lambda\Lambda+2N\Xi-\sqrt{3}\Sigma\Sigma)$ & $-2$  \\
3 & [3,2,1] $\bf 8$ & 1/2 & 2$^+$ & $\frac{1}{\sqrt{5}}
(\sqrt{2}N\Omega-\Lambda\Xi^{\ast}+\Sigma^{\ast}\Xi-\Sigma\Xi^{\ast})$&$-1$ \\
\hline
\end{tabular}
\end{center}
\end{table} 

Except for ${\cal S}=-1$, the leading dibaryon candidates listed in 
Table~\ref{tab:oka} are the ones mostly dealt with in quark-model 
calculations. The table shows clearly the prominence of the ${\cal S}=-2$ $H$ 
dibaryon that was first predicted by Jaffe \cite{jaffe77} as a genuine bound 
state well below the $\Lambda\Lambda$ threshold. However, more realistic $6q$ 
quark cluster model calculations that (i) break SU(3)$_{\rm f}$, (ii) account 
for full quark antisymmetrization, and (iii) also make contact via resonating 
group methods (RGM) with related $BB^\prime$ coupled channels and thresholds, 
placed the $H$ near the $\Xi N$ threshold at $E_{\Lambda\Lambda}\approx 
26$~MeV \cite{oka83}. Recent experimental searches for a weakly decaying 
$\Lambda\Lambda$ bound state by Belle \cite{Belle13} and ALICE \cite{ALICE15} 
imply that Jaffe's $H$ dibaryon is particle-unstable against strong decay. 
This is confirmed by recent lattice QCD (LQCD) simulations \cite{HALQCD12} 
and by chiral EFT arguments \cite{haidenbauer12} suggesting that the $H$ 
could appear at most as a resonance near the $\Xi N$ threshold at $E_{\Lambda
\Lambda}\approx 26$~MeV, in agreement with the prediction of the 1983 first 
$6q$ RGM calculation \cite{oka83}. For ${\cal S}=-3$, the $2^+$ deeply bound 
$\Omega N$ dibaryon predicted in Ref.~\cite{goldman87}, together with a $1^+$ 
companion, is more likely according to recent LQCD simulations \cite{HALQCD14} 
to be just weakly bound with respect to the $\Omega$--$N$ threshold, well 
above the lower ${\cal S}=-3$ thresholds $\Xi$--$\Lambda$ and $\Xi$--$\Sigma$, 
again far from being particle-stable. 

For ${\cal S}=0$, although the recently established ${\cal D}_{03}(2380)$ 
\cite{wasa11} lies below the $\Delta\Delta$ threshold, it is far from 
being particle-stable and is considerably less bound than suggested e.g. 
in Ref.~\cite{goldman89}. In fact, a recent study of non-strange $6q$ 
spatially symmetric $L=0$ dibaryons \cite{PPL15}, superseding $6q$ bag-model 
calculations \cite{jaffe77,MT83}, finds such a $\Delta\Delta$ 
dibaryon at several hundreds of MeV above the $\Delta$--$\Delta$ threshold, 
concluding that ``the recently observed peak in the $I(J^P)$=$0(3^+)$ channel 
should be a molecular configuration composed of two $\Delta$ baryons." Indeed, 
the hadronic-based calculations reviewed below emphasize the long-range 
physics aspects of non-strange dibaryons.

\section{Non-strange dibaryons} 
\label{sec:S=0}

$N\Delta$ and $\Delta\Delta$ $s$-wave dibaryon resonances 
${\cal D}_{IS}$ with isospin $I$ and spin $S$ were proposed as early as 
1964, when quarks were still perceived as merely mathematical entities, 
by Dyson and Xuong \cite{dyson64} who focused on the lowest-dimension 
SU(6) multiplet in the $\bf{56\times 56}$ product that contains the SU(3) 
$\overline{\bf 10}$ and ${\bf 27}$ multiplets in which the deuteron 
${\cal D}_{01}$ and $NN$ virtual state ${\cal D}_{10}$ are classified. 
This yields two dibaryon candidates, ${\cal D}_{12}$ ($N\Delta$) and 
${\cal D}_{03}$ ($\Delta\Delta$) as listed in Table~\ref{tab:dyson}. 
Identifying the constant $A$ in the resulting mass formula 
$M=A+B[I(I+1)+S(S+1)-2]$ with the $NN$ threshold mass 1878~MeV, 
a value $B\approx 47$~MeV was determined by assigning ${\cal D}_{12}$ 
to the $pp\leftrightarrow \pi^+ d$ resonance at $\sqrt{s}=2160$~MeV 
(near the $N\Delta$ threshold) which was observed already during 
the 1950's. This led to the prediction $M({\cal D}_{03})$=2350~MeV. 
The ${\cal D}_{03}$ dibaryon was the subject of many quark-based 
model calculations since 1980, as reviewed elsewhere \cite{gg14}. 

\begin{table}[hbt]
\begin{center} 
\caption{Non-strange $s$-wave dibaryon SU(6) predictions \cite{dyson64}. 
The ${\overline{\bf 10}}$ SU(3)$_{\rm f}$ representation is denoted here 
${\bf 10}^{\ast}$.} 
\begin{tabular}{cccccc}
\hline
dibaryon & $I$ & $S$ & SU(3) & legend & mass \\
\hline 
${\cal D}_{01}$ & 0 & 1 & ${\bf 10}^{\ast}$ & deuteron & $A$ \\ 
${\cal D}_{10}$ & 1 & 0 & ${\bf 27}$ & $nn$ & $A$ \\ 
${\cal D}_{12}$ & 1 & 2 & ${\bf 27}$ & $N\Delta$ & $A+6B$ \\ 
${\cal D}_{21}$ & 2 & 1 & ${\bf 35}$ & $N\Delta$ & $A+6B$ \\ 
${\cal D}_{03}$ & 0 & 3 & ${\bf 10}^{\ast}$ & $\Delta\Delta$ & $A+10B$ \\ 
${\cal D}_{30}$ & 3 & 0 & ${\bf 28}$ & $\Delta\Delta$ & $A+10B$ \\ 
\hline 
\end{tabular} 
\label{tab:dyson}
\end{center} 
\end{table} 

It is shown below that the pion-assisted methodology applied recently 
by Gal and Garcilazo \cite{galgar13,galgar14} couples ${\cal D}_{12}$ and 
${\cal D}_{03}$ dynamically in a perfectly natural way, the analogue of which 
has not emerged in quark-based models. As stated earlier in this Review, 
our hadronic-based calculations emphasize the long-range physics aspects of 
non-strange dibaryons.

\subsection{$N\Delta$ dibaryons} 

\begin{figure}[htb] 
\begin{center} 
\includegraphics[width=0.48\textwidth,height=6cm]{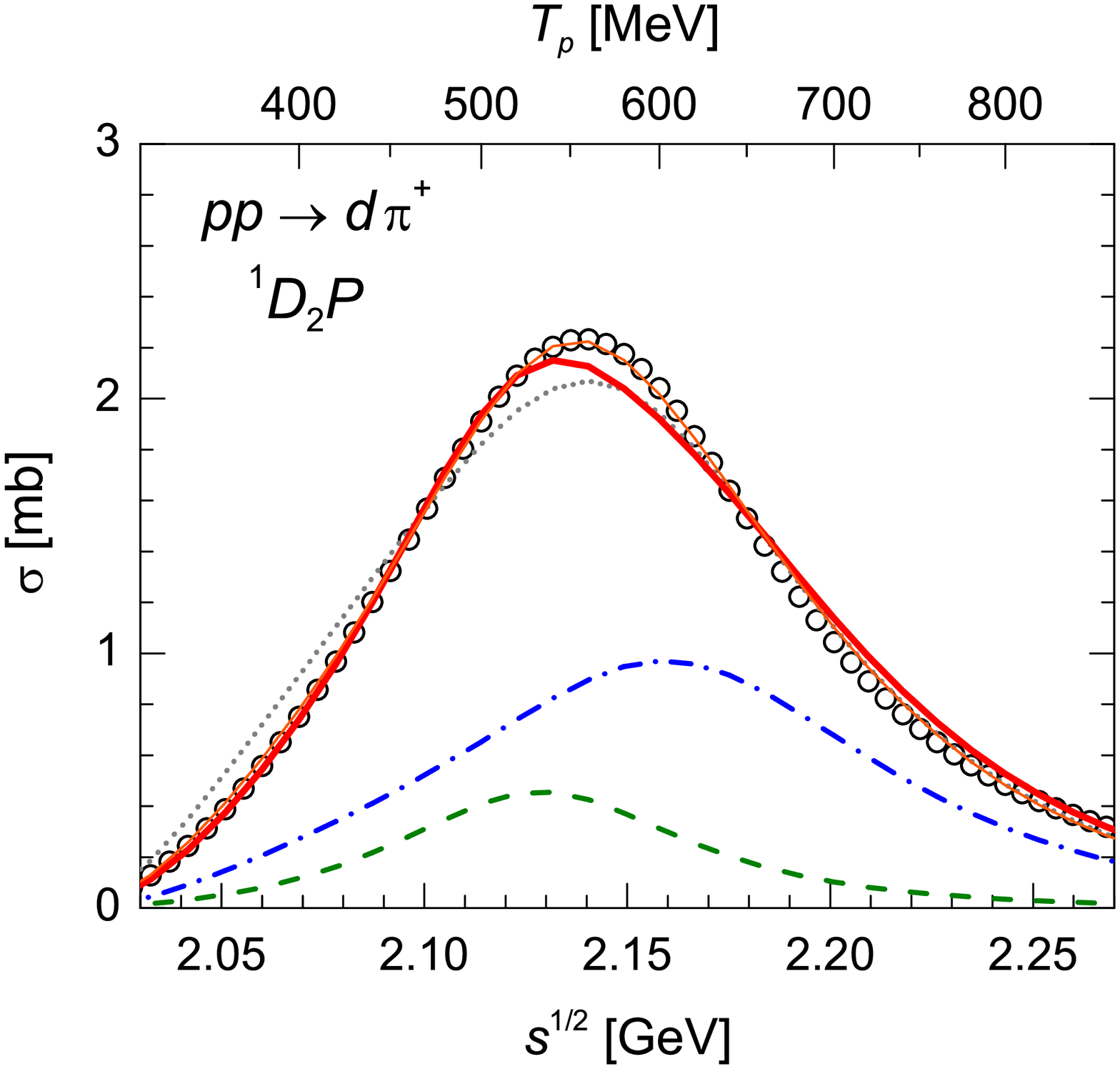} 
\includegraphics[width=0.48\textwidth,height=6cm]{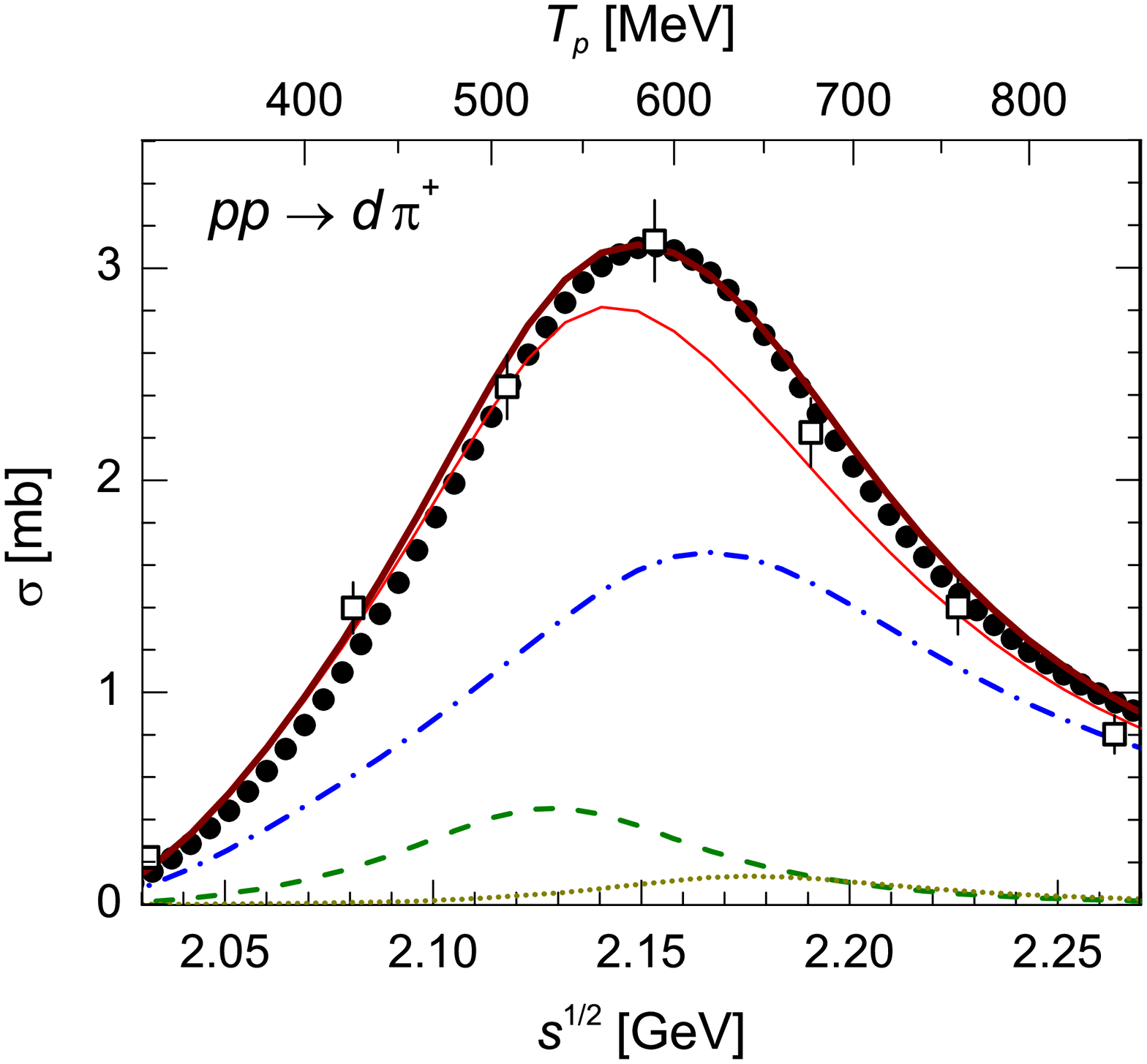} 
\caption{${\cal D}_{12}$ dibaryon $s$-channel (dashed) contributions to $pp\to 
d\pi^+$ ${^1D_2}{^3P_2}$ partial-wave (left panel) and total (right panel) 
cross sections from SAID \cite{SAID}, plus a small ${^3F_3}{^3D_3}$ dibaryon 
(dotted) contribution, in a model \cite{PK15} that includes non-resonant 
$t$-channel exchange (dot-dashed) contributions with amplitudes interfering 
constructively with $s$-channel amplitudes. Model sensitivities are exhibited 
in thin lines.} 
\label{fig:D12PK} 
\end{center} 
\end{figure} 

The ${\cal D}_{12}$ dibaryon shows up experimentally as $NN({^1D_2})$ 
$\leftrightarrow$ $\pi d({^3P_2})$ coupled-channel resonance corresponding 
to a quasibound $N\Delta$ with mass $M\approx 2.15$~GeV, near the $N\Delta$ 
threshold, and width $\Gamma\approx 0.12$~GeV as derived from the Argand 
diagram of the $^1D_2$ partial wave in $pp$ elastic scattering, using the 
SAID partial-wave analysis \cite{SAID}. The contribution of ${\cal D}_{12}$ 
to the $pp\to d\pi^+$ cross section in a recent reaction model calculation 
\cite{PK15} is shown by dashed lines in Fig.~\ref{fig:D12PK}. 

\begin{figure}[hbt] 
\begin{center} 
\includegraphics[width=0.7\textwidth]{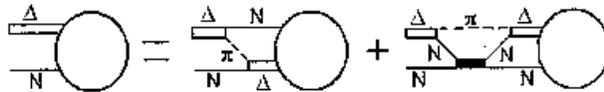} 
\end{center} 
\caption{$N\Delta$ dibaryon's Lippmann-Schwinger equation \cite{galgar14}.} 
\label{fig:DIS} 
\end{figure}

In our recent work \cite{galgar14} we have calculated this dibaryon 
and other $N\Delta$ dibaryon candidates such as ${\cal D}_{21}$ 
(see Table~\ref{tab:dyson}) by solving Faddeev equations with relativistic 
kinematics for the $\pi NN$ three-body system, where the $\pi N$ subsystem 
is dominated by the $P_{33}$ $\Delta$(1232) resonance channel and the $NN$ 
subsystem is dominated by the $^3S_1$ and $^1S_0$ channels. The coupled 
Faddeev equations give rise then to an effective $N\Delta$ Lippmann-Schwinger 
(LS) equation for the three-body $S$-matrix pole, with energy-dependent 
kernels that incorporate spectator-hadron propagators, as shown 
diagrammatically in Fig.~\ref{fig:DIS} where circles denote the $N\Delta$ 
$T$ matrix. 

Of the four $L=0$, $IS=12,21,11,22$ $N\Delta$ dibaryon candidates 
${\cal D}_{IS}$, the latter two do not provide resonant solutions. 
For ${\cal D}_{12}$ (${\cal D}_{21}$), only $^3S_1$ ($^1S_0$) contributes 
out of the two $NN$ interactions. Since the $^3S_1$ interaction is the more 
attractive one, ${\cal D}_{12}$ lies below ${\cal D}_{21}$ as borne out by 
the calculated masses listed in Table~\ref{tab:NDel} for two choices of the 
$P_{33}$ interaction form factor corresponding to $\Delta$-isobar spatial 
sizes 1.35 and 0.9~fm. The two dibaryons are found to be degenerate to 
within less than 20~MeV. The mass values calculated for ${\cal D}_{12}$ 
are reasonably close to those from Refs.~\cite{arndt87,hosh92}. 

\begin{table}[hbt] 
\begin{center} 
\caption{$N\Delta$ dibaryon $S$-matrix poles (in MeV) for ${\cal D}_{12}$ 
and ${\cal D}_{21}$ obtained by solving the LS equation, Fig.~\ref{fig:DIS}, 
derived from $\pi NN$ Faddeev equations \cite{galgar14} are listed for 
large ($>$) and small ($<$) sized $\pi N$ $P_{33}$ form factors and also 
cited from non-Faddeev determinations \cite{arndt87,hosh92}.} 
\begin{tabular}{ccccccc} 
\hline  
${\cal D}_{12}(>)$ & ${\cal D}_{21}(>)$ & ${\cal D}_{12}(<)$ & 
${\cal D}_{21}(<)$ &  & ${\cal D}_{12}$~\cite{arndt87} & 
${\cal D}_{12}$~\cite{hosh92}  \\  
\hline   
2147$-{\rm i}$60 & 2165$-{\rm i}$64 &  2159$-{\rm i}$70 & 
2169$-{\rm i}$69 &  & 2148$-{\rm i}$63 & 2144$-{\rm i}$55   \\
\hline 
\end{tabular} 
\label{tab:NDel} 
\end{center} 
\end{table}

\subsection{$\Delta\Delta$ dibaryons} 

\begin{figure}[htb] 
\begin{center} 
\includegraphics[width=0.48\textwidth,height=5.5cm]{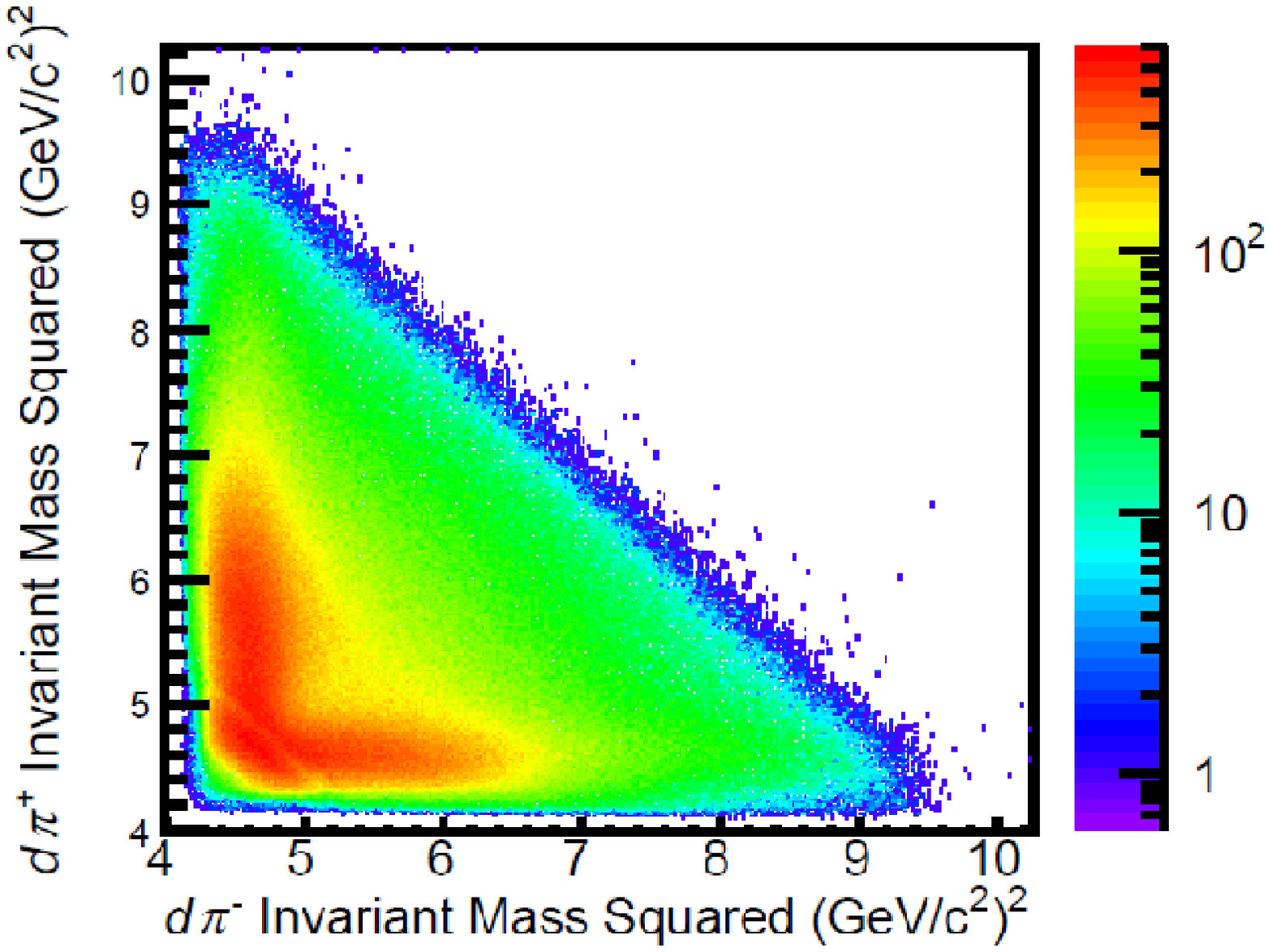} 
\includegraphics[width=0.48\textwidth,height=5.5cm]{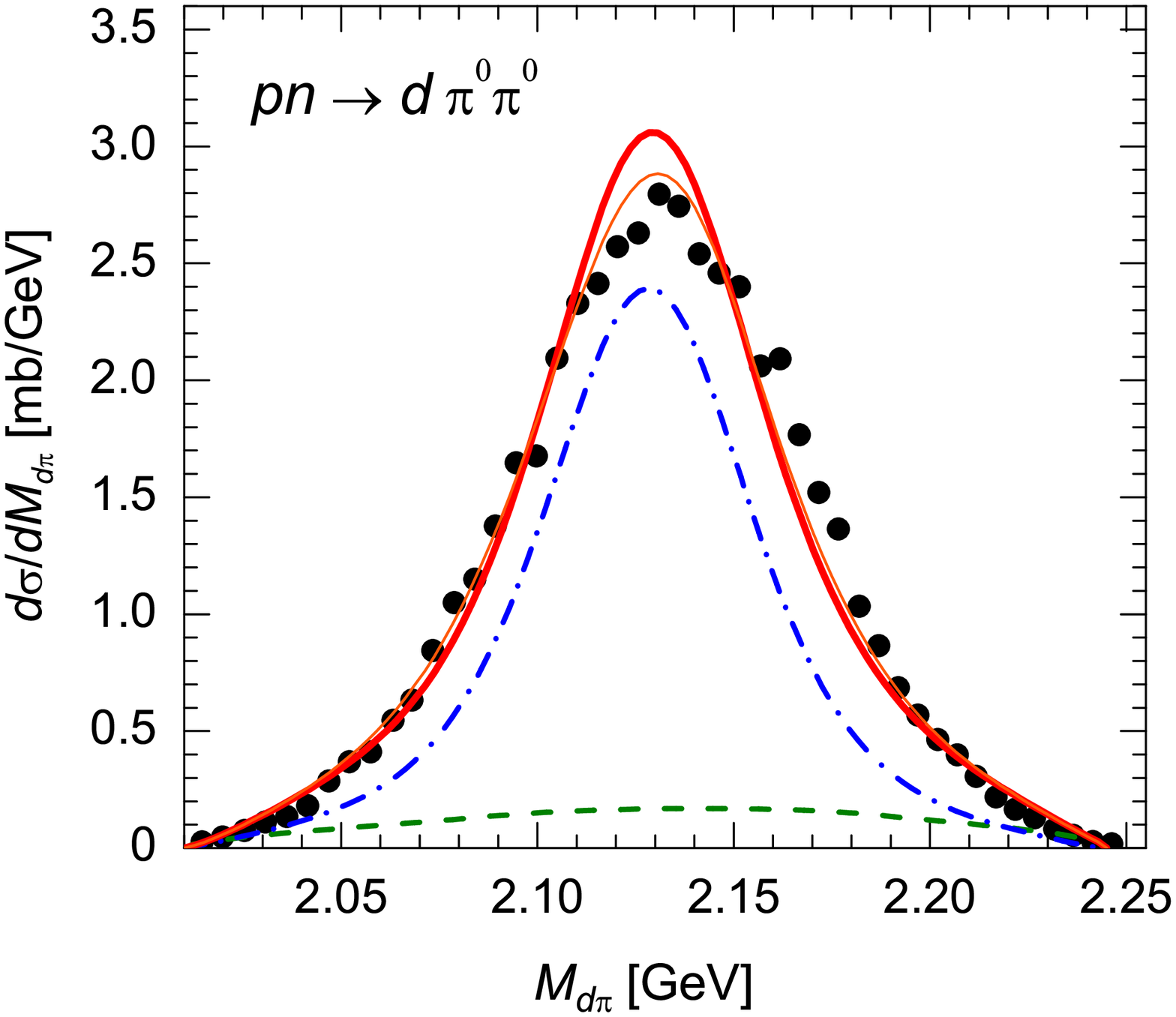} 
\caption{Left: ${\cal D}_{12}(2150)$ $N\Delta$ dibaryon resonance signal 
in the Dalitz plot of $M^2_{d\pi^+}$ {\it vs} $M^2_{d\pi^-}$ from preliminary 
$\gamma d \to d \pi^+ \pi^-$ measurements by the CLAS g13 Collaboration at 
JLab \cite{APS15}. Right: WASA-at-COSY $M_{d\pi}$ distribution \cite{wasa11} 
and as calculated for two (solid lines) input parametrizations of 
${\cal D}_{12}(2150)$ \cite{PK15}. The dot-dashed line gives the 
${\cal D}_{12}(2150)+\pi$ contribution to the two-body decay of 
${\cal D}_{03}(2380)$, and the dashed line gives a scalar-isoscalar 
emission contribution.}  
\label{fig:D03PK} 
\end{center} 
\end{figure} 

The relevance of the ${\cal D}_{12}(2150)$ $N\Delta$ dibaryon to the physics 
of the ${\cal D}_{03}(2380)$ $\Delta\Delta$ dibaryon is demonstrated 
in Fig.~\ref{fig:D03PK} by showing, on the left panel, a $d\pi^{\pm}$ 
invariant-mass correlation near the $N\Delta$ threshold as deduced from 
preliminary CLAS data on the $\gamma d\to d\pi^+\pi^-$ reaction \cite{APS15} 
and, on the right panel, a $d\pi$ invariant-mass distribution peaking near 
the $N\Delta$ threshold as deduced from the WASA-at-COSY $pn\to d\pi^0\pi^0$ 
reaction by which the ${\cal D}_{03}(2380)$ dibaryon was discovered 
\cite{wasa11}. The $\gamma d\to d\pi^+\pi^-$ preliminary CLAS data suggest 
a subthreshold ${\cal D}_{12}(2150)$ dibaryon with mass 2115$\pm$10~MeV 
and width 125$\pm$25~MeV, consistently with past deductions. The peaking of 
the $d\pi$ invariant-mass distribution in the $pn\to d\pi^0\pi^0$ reaction 
essentially at this ${\cal D}_{12}(2150)$ mass value suggests that the 
two-body decay modes of ${\cal D}_{03}(2380)$ are almost saturated 
by the ${\cal D}_{12}(2150)+\pi$ decay mode, as reflected in the 
calculation \cite{PK15} depicted in the right panel. 

Four-body $\pi\pi NN$ calculations are required, strictly speaking, 
to discuss $\Delta\Delta$ dibaryons. In Ref.~\cite{galgar13} we studied the 
${\cal D}_{03}$ dibaryon by solving a $\pi N\Delta'$ three-body model, 
where $\Delta'$ is a stable $\Delta$(1232) and the $N\Delta'$ interaction 
is dominated by the ${\cal D}_{12}$ dibaryon. The $I(J^P)=1(2^+)$ 
$N\Delta'$ interaction was not assumed to resonate, but was fitted within 
a $NN$--$\pi NN$--$N\Delta'$ coupled-channel caricature model to the 
$NN$ $^1D_2$ $T$-matrix, requiring that the resulting $N\Delta'$ 
separable-interaction form factor is representative of long-range 
physics, with momentum-space soft cutoff $\Lambda$ below 3.5~fm$^{-1}$. 

\begin{figure}[htb] 
\begin{center} 
\includegraphics[width=0.6\textwidth]{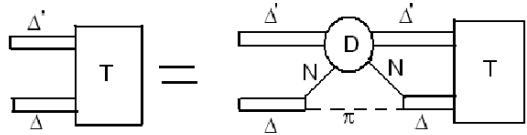} 
\caption{$S$-matrix pole equation for ${\cal D}_{03}(2370)$ $\Delta\Delta$ 
dibaryon \cite{galgar13}.} 
\label{fig:D03} 
\end{center} 
\end{figure} 

The Faddeev equations of the $\pi N\Delta'$ three-body model give rise, 
as before, to an effective LS equation for the $\Delta\Delta'$ $S$-matrix pole 
corresponding to ${\cal D}_{03}$. This LS equation is shown diagrammatically 
in Fig.~\ref{fig:D03}, where $D$ stands for the ${\cal D}_{12}$ dibaryon. 
The $\pi N$ interaction was assumed again to be dominated by the $P_{33}$ 
$\Delta$ resonance, using two different parametrizations of its form factor 
that span a reasonable range of the $\Delta$ hadronic size. 
In Ref.~\cite{galgar14} we have extended the calculation of ${\cal D}_{03}$ 
to other ${\cal D}_{IS}$ $\Delta\Delta$ dibaryon candidates, with $D$ now 
standing for both $N\Delta$ dibaryons ${\cal D}_{12}$ and ${\cal D}_{21}$. 
Since ${\cal D}_{21}$ is almost degenerate with ${\cal D}_{12}$, and 
with no $NN$ observables to constrain the input $(I,S)$=(2,1) $N\Delta'$ 
interaction, the latter was taken the same as for $(I,S)$=(1,2). The model 
dependence of this assumption requires further study. ${\cal D}_{03}$ and 
${\cal D}_{30}$ are the lowest and narrowest $\Delta\Delta$ dibaryons. 

\begin{table}[hbt] 
\begin{center} 
\caption{$\Delta\Delta$ dibaryon $S$-matrix poles (in MeV) obtained in 
Refs.~\cite{galgar13,galgar14} by using a spectator-$\Delta'$ complex mass 
$W(\Delta')$ (first column) in the propagator of the LS equation depicted 
in Fig.~\ref{fig:D03}. The superscripts $>$ and $<$ stand for two choices 
of the $\pi N$ $P_{33}$ form factor, with spatial sizes of 1.35~fm ($>$) 
and 0.9~fm ($<$).}  
\begin{tabular}{ccccc} 
\hline  
$W(\Delta')$ & $W^{>}({\cal D}_{03})$ & $W^{>}({\cal D}_{30})$ & 
$W^{<}({\cal D}_{03})$ & $W^{<}({\cal D}_{30})$  \\  
\hline
1211$-{\rm i}$49.5 & 2383$-{\rm i}$47 & 2412$-{\rm i}$49 & 
2342$-{\rm i}$31 & 2370$-{\rm i}$30  \\
1211$-{\rm i}$(2/3)49.5 & 2383$-{\rm i}$41 & 2411$-{\rm i}$41 & 
2343$-{\rm i}$24 & 2370$-{\rm i}$22  \\
\hline  
\end{tabular} 
\label{tab:DelDel} 
\end{center} 
\end{table} 

Representative results for ${\cal D}_{03}$ and ${\cal D}_{30}$ are assembled 
in Table~\ref{tab:DelDel}, where the calculated mass and width values listed 
in each row correspond to the value listed there of the spectator-$\Delta'$ 
complex mass $W(\Delta')$ used in the propagator of the LS equation shown in 
Fig.~\ref{fig:D03}. The value of $W(\Delta')$ in the first row is that of the 
$\Delta$(1232) $S$-matrix pole. It is implicitly assumed thereby that the 
decay $\Delta' \to N\pi$ proceeds independently of the $\Delta \to N\pi$ 
isobar decay. However, as pointed out in Ref.~\cite{galgar13}, care must be 
exercised to ensure that the decay nucleons and pions satisfy Fermi-Dirac 
and Bose-Einstein statistics requirements, respectively. Assuming $L=0$ for 
the decay-nucleon pair, this leads to the suppression factor 2/3 depicted 
in the value of $W(\Delta')$ listed in the second row. It is seen that the 
widths obtained upon applying this width-suppression are only moderately 
smaller, by less than 15 MeV, than those calculated disregarding this 
quantum-statistics correlation. A more complete discussion of these and 
of other ${\cal D}_{IS}$ $\Delta\Delta$ dibaryon candidates is found in 
Ref.~\cite{galgar14}. 

The mass and width values $W^{>}({\cal D}_{03})$ in Table~\ref{tab:DelDel} 
agree very well with those determined by the WASA-at-COSY Collaboration 
\cite{wasa11,wasa13,wasa14}, reproducing in particular the reported width 
$\Gamma({\cal D}_{03})\approx 80$~MeV which is considerably below the rough 
estimate $2\Gamma_{\Delta}\approx 200$~MeV for two free-space $\Delta$'s, 
using the $\Delta(1232)$ pole position from SAID \cite{SAID}. However, 
the reduced phase space for each $\Delta\to N\pi$ decay suppresses this 
estimate by a factor 0.555, which together with the suppression factor 2/3 
from the previous paragraph yields the estimate $\Gamma(\Delta\Delta)_{03}
\approx 73$~MeV, to which the partial decay widths to $NN\pi$ and $NN$ 
need to be added. This results in a total width estimate of about 90~MeV, 
compared to 82~MeV from Table~\ref{tab:DelDel}. A similar estimate can be 
obtained by considering ${\cal D}_{03}$ decay as occurring through its lower 
$\pi{\cal D}_{12}$ channel. 

The ${\cal D}_{30}$ dibaryon in our calculations is located only 
$\approx$30~MeV above ${\cal D}_{03}$, and with a similar width. 
Allowing its ${\cal D}_{21}$ input parameters to depart from those 
found for ${\cal D}_{12}$ would increase the ${\cal D}_{30}$ mass 
by 20--30~MeV, in close agreement with the quark-based calculations 
of Ref.~\cite{HPW14}. Note, however, that the widths calculated there 
are much larger than ours. The $I=3$ exotic ${\cal D}_{30}$ dibaryon 
was discussed in Ref.~\cite{BBC13}, where the dominant role that six-quark 
hidden-color (HC) configurations might play in binding ${\cal D}_{03}$ 
and ${\cal D}_{30}$ was emphasized. However, recent explicit quark-based 
calculations \cite{HPW14} find HC configurations to play a marginal role, 
enhancing dibaryon binding by merely 15$\pm$5~MeV and reducing the dibaryon 
width from 175 to 150 MeV for ${\cal D}_{03}$, still twice as big as the 
reported width, and from 216 to 200 MeV for ${\cal D}_{30}$. This is in line 
with the negligible role found long ago for HC configurations in the dibaryon 
calculation of Ref.~\cite{oka82}. In contrast, a very recent calculation 
\cite{dong15} claims that $6q$ HC configurations reduce substantially the 
calculated width of ${\cal D}_{03}$ down to $\Gamma\approx 70$~MeV, the 
argument given being that HC components cannot decay to colorless hadrons. 
This argument overlooks the strong coupling between colorless and HC $BB'$ 
components in any realistic $6q$ wavefunction, through which the HC components 
decay by using the colorless components for intermediate states.

\section{Strange dibaryons} 
\label{sec:S=-1} 

\begin{figure}[htb] 
\begin{center} 
\includegraphics[width=0.6\textwidth]{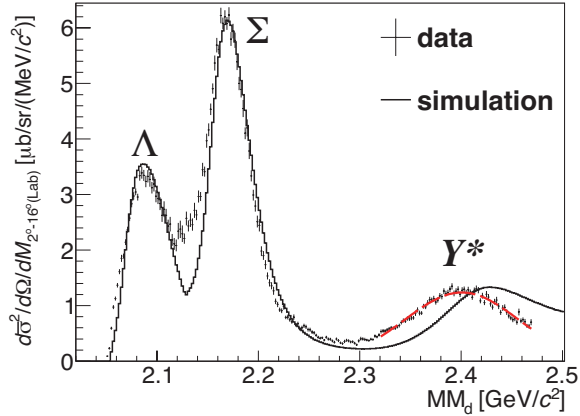}
\caption{J-PARC E27 missing-mass spectrum in $d(\pi^+,K^+)$ 
at 1.69 GeV/c \cite{E27a}} 
\label{fig:Y*N} 
\end{center} 
\end{figure}

Recent searches for a ${\bar K}NN$ (known as $K^-pp$) 
$I(J^P)$=$\frac{1}{2}(0^-)$ dibaryon have been reported by experiments 
at Frascati \cite{finuda13}, SPring-8 \cite{leps14}, GSI \cite{gsi14a,gsi15} 
and J-PARC \cite{E27a,E27b,E15a,E15b}. A missing-mass spectrum measured 
in the $d(\pi^+,K^+)$ reaction at 1.69 GeV/c in J-PARC is shown in 
Fig.~\ref{fig:Y*N}, indicating $\approx$22~MeV attractive shift of 
the unresolved $Y^{\ast}(1385+1405)$ quasi-free peak complex. This is 
consistent with the attraction expected in the $I(J^P)$=$\frac{1}{2}(0^-)$ 
$\Lambda(1405)N$ $s$-wave channel shown in Ref.~\cite{oka11} to overlap 
substantially with $K^-pp$. Chirally motivated calculations of $K^-pp$ find 
binding energies of few tens of MeV and larger widths, see the recent review 
\cite{gal13}. Such relatively shallow $K^-pp$ binding persists upon including 
the $\pi\Lambda N$ and $\pi\Sigma N$ lower-mass channels~\cite{RS14}. There is 
a hint of a very broad bound-state signal about 15~MeV below threshold from 
J-PARC experiment E15 \cite{E15b}, while several past experiments, notably 
the recent J-PARC experiment E27 \cite{E27b}, claimed a bound state signal, 
also very broad, near the $\pi\Sigma N$ threshold about 100~MeV below the 
$K^-pp$ threshold. Such a deeply bound $I(J^P)$=$\frac{1}{2}(0^-)$ $K^-pp$ 
state is unacceptable theoretically. 

The $\pi\Lambda N$--$\pi\Sigma N$ system, however, may benefit from strong 
meson-baryon $p$-wave interactions, fitted to the $\Delta(1232)\to\pi N$ and 
$\Sigma(1385)\to\pi\Lambda$--$\pi\Sigma$ form factors, by aligning isospin 
and angular momentum to $I(J^P)$=$\frac{3}{2}(2^+)$. Such a ${\cal S}=-1$ 
pion assisted dibaryon was studied in Ref.~\cite{gg13} by solving $\pi YN$ 
coupled-channel Faddeev equations, thereby predicting a dibaryon resonance 
${\cal Y}_{\frac{3}{2}2^+}$ slightly below the $\pi\Sigma N$ threshold 
($\sqrt{s_{\rm th}}\approx 2270$~MeV). Adding a $\bar KNN$ channel hardly 
matters, since its leading $^3S_1$ $NN$ configuration is Pauli forbidden. 
Note that with isospin $I=\frac{3}{2}$, this dibaryon differs from the 
$I(J^P)$=$\frac{1}{2}(0^-)$ $K^-pp$ and from the $I(J^P)$=$\frac{1}{2}(2^+)$ 
dibaryon listed in Table~\ref{tab:oka} which according to our calculations 
might lie almost 100~MeV above ${\cal Y}_{\frac{3}{2}2^+}(2270)$. 

The ${\cal S}=-1$ ${\cal Y}_{\frac{3}{2}2^+}(2270)$ dibaryon is expected 
to have good overlap with $^5S_2$, $I=\frac{3}{2}$ $\Sigma(1385)N$ and 
$\Delta(1232)Y$ dibaryon configurations, the lower of which $\Sigma(1385)N$ 
lies about 50~MeV above the $\pi\Sigma N$ threshold. We emphasize that these 
quantum numbers differ from $^1S_0$, $I=\frac{1}{2}$ for $\Lambda(1405)N$ 
which is normally being searched upon. A recent search in 
\begin{eqnarray} 
   p~ + ~p & ~\rightarrow ~ & {\cal Y}^{++} ~+~K^0 \nonumber  \\  
           &                & ~\hookrightarrow ~ \Sigma^+ ~+~ p \; 
\label{eq:pptoY++} 
\end{eqnarray} 
by the HADES Collaboration at GSI \cite{gsi14b} found no $\cal Y$ dibaryon 
signal. It is not clear whether the $pp$ experiments were able to deal with 
as small cross sections as 0.1 $\mu$b or less that are likely to be needed 
in order to excite $\cal Y$ dibaryon candidates \cite{gsi15}. Other possible 
search reactions are  
\begin{eqnarray} 
\pi^{\pm} ~+~ d & ~\rightarrow ~ & {\cal Y}^{++/-} ~+~K^{0/+}  \nonumber  \\ 
 &  &  ~\hookrightarrow ~ \Sigma^{\pm} +p(n) \; , 
\label{eq:pi+dtoY++} 
\end{eqnarray} 
again offering distinct $I=\frac{3}{2}$ decay channels. Other decay channels 
such as 
\begin{eqnarray} 
\pi^+ ~+~ d & ~\rightarrow ~ & {\cal Y}^+ ~+~K^+  \nonumber  \\ 
 &  &  ~\hookrightarrow ~ \Sigma^0 + p  
\label{eq:pi+dtoY+} 
\end{eqnarray} 
allow for both $I=\frac{1}{2},\frac{3}{2}$. E27 has just reported \cite{E27b} 
a dibaryon signal near the $\pi\Sigma N$ threshold in reaction 
(\ref{eq:pi+dtoY+}). This requires further experimental study.

\section{Charmed dibaryons} 
\label{sec:charm} 

Pion assisted dibaryon candidates in the charm ${\cal C}=+1$ sector have been 
discussed recently in Ref.~\cite{ggvc14}. In this work the same formalism 
applied earlier in the strangeness ${\cal S}=-1$ sector to the $\pi\Lambda 
N$ system \cite{gg13} was applied to the charmed $\pi\Lambda_c N$ system, 
replacing the $\Lambda(1116)$ baryon by the $\Lambda_c(2286)$ charmed baryon 
and the $\Sigma(1385)$ resonance by the $\Sigma_c(2520)$ charmed resonance,  
but disregarding the coupling of $\pi\Lambda_c(2286)N$ to $\pi\Sigma(2455)N$. 
The $\Lambda_c(2286)N$ system was studied in a chiral constituent quark model 
\cite{valcarce05} with a separable $s$-wave interaction. Separable $p$-wave 
interactions were used for the pion-baryon channels, dominated here by the 
$\Delta(1232)$ and $\Sigma_c(2520)$ resonances. Faddeev equations using 
relativistic kinematics were solved to look for bound states and resonances 
with quantum numbers $I(J^P)$=$\frac{3}{2}(2^+)$. Some of the tested models 
generated a very narrow bound-state or resonance below the $\Sigma_c(2455)N$ 
threshold, violating isospin in its strong decay to $\Lambda_c(2286)N$. 
Note that the $\Sigma_c(2455)N$ threshold lies $\approx$27~MeV above the 
$\pi\Lambda_c(2286)N$ threshold. The prediction of this charmed pion assisted 
dibaryon is robust since it depends little on the $\Lambda_c N$ spin-triplet 
$s$-wave interaction, even if the precise energy of the resonance is 
not pinned down between threshold at $\sqrt{s_{\rm th}}\approx 3363$~MeV 
and several tens of MeV above threshold. This resonance may be viewed 
as a $\Sigma_c(2520)N$ dibaryon bound state and is likely to be the 
{\it lowest lying} charmed dibaryon, considerably below the mass 
$\approx$3500~MeV predicted recently for a $DNN$ bound state with 
quantum numbers $I(J^P)$=$\frac{1}{2}(0^-)$ that may be viewed also as a 
$\Lambda_c(2595)N$ dibaryon bound state \cite{DNN12}. The $DNN$ bound state 
resembles in structure and quantum numbers the $K^-pp$ quasibound state that 
may also be viewed as a $\Lambda(1405)N$ dibaryon bound state. 

Denoting the $I(J^P)=\frac{3}{2}(2^+)$ $\pi\Lambda_c N$ dibaryon by 
${\cal C}$, this ${\cal C}_{\frac{3}{2}2^+}(3370)$ dibaryon candidate 
could be searched with proton and pion beams in the high-momentum 
hadron beam line extension approved at J-PARC by, e.g. 
\begin{eqnarray} 
   p~ + ~p & ~\rightarrow ~ & {\cal C}^{+++} ~+~D^- \nonumber  \\  
           &                & ~\hookrightarrow ~ \Sigma_c^{++}(2455)~+~ p \; ,
\label{eq:pptoYc+++} 
\end{eqnarray} 
\begin{eqnarray} 
\pi^+ ~+~ d & ~\rightarrow ~ & {\cal C}^{+++} ~+~D^-  \nonumber  \\ 
 &  &  ~\hookrightarrow ~ \Sigma_c^{++}(2455)~+~ p \; , 
\label{eq:pi+dtoYc+++}  
\end{eqnarray} 
\begin{eqnarray} 
\pi^- ~+~ d & ~\rightarrow ~ & {\cal C}^+ ~+~D^-  \nonumber  \\ 
 &  &  ~\hookrightarrow ~ \Sigma_c^{+/0}(2455)~+~ n/p \; .  
\label{eq:pi-dtoYc+}  
\end{eqnarray} 
The ${\cal C}_{\frac{3}{2}2^+}(3370)$ dibaryon may be looked for both within 
inclusive missing-mass measurements that focus on the outgoing $D^-$ charmed 
meson, and in exclusive invariant-mass measurements that focus on the decay 
$\Sigma_c(2455)N$ pair, provided that ${\cal C}$ is located above the 
$\Sigma_c(2455)N$ threshold.

\section{Conclusion} 

It was shown how the 1964 Dyson-Xuong SU(6)-based classification and 
predictions of non-strange dibaryons \cite{dyson64} are confirmed in the 
hadronic model of $N\Delta$ and $\Delta\Delta$ pion-assisted dibaryons 
\cite{galgar13,galgar14}. The input for dibaryon calculations in this model 
consists of nucleons, pions and $\Delta$'s, interacting via long-range 
pairwise interactions. These calculations reproduce the two nonstrange 
dibaryons established experimentally and phenomenologically so far, the 
$N\Delta$ dibaryon ${\cal D}_{12}$ \cite{arndt87,hosh92} and the $\Delta
\Delta$ dibaryon ${\cal D}_{03}$ \cite{wasa11,wasa13,wasa14}, and predict 
several exotic $N\Delta$ and $\Delta\Delta$ dibaryons. We note that, within 
the $\pi N\Delta$ three-body model of ${\cal D}_{03}$, ${\cal D}_{12}$ 
provides a two-body decay channel $\pi {\cal D}_{12}$ with threshold lower 
than $\Delta\Delta$ which proves instrumental in obtaining a relatively small 
width for ${\cal D}_{03}$ \cite{galgar14}.  

Finally, straightforward extensions of $\cal S$=0 pion-assisted dibaryon 
phenomenology to strangeness $\cal S$=$-1$ and charm $\cal C$=$+1$ were 
briefly discussed, mostly in relation to recent searches of kaonic 
nuclear clusters~\cite{gal13}.

\section*{Acknowledgments} 
Stimulating discussions with Mikhail Bashkanov and Heinz Clement on 
dibaryons, as well as the kind hospitality by Pawel Moskal and his 
group at the Jagiellonian Symposium on Fundamental and Applied 
Subatomic Physics, Krakow, June 2015, are gratefully acknowledged.

\end{document}